\documentclass[ twocolumn]{IEEEtran}
\usepackage{url,amsmath,graphicx}
 \usepackage{stfloats}
 \usepackage{eqparbox}
  \usepackage{array}
  \usepackage{fixltx2e}
  \usepackage{stfloats}
   \usepackage{cite}
   \usepackage{url}
\usepackage{fancyhdr}
\usepackage{cite}
\usepackage{graphicx}
\usepackage{psfrag}
\usepackage{url}
\usepackage{stfloats}
\usepackage{amsmath}
\usepackage{array}
\usepackage{fancyhdr}

\usepackage{amssymb}
\usepackage{color}

\begin{document}
\title{ Coverage Analysis and  Scaling  Laws \\in  Ultra-Dense  Networks}

\author{ Im\`ene~Trigui, Member,  \textit{IEEE},  Sofi\`ene~Affes, Senior Member, \textit{IEEE}, Marco~Di~Renzo, Fellow, \textit{IEEE},   and Dushantha~Nalin~K.~Jayakody, Senior Member, \textit{IEEE}}
%

 \maketitle

\begin{abstract}

In this  paper, we develop an innovative approach to  quantitatively characterize the performance
of ultra-dense  wireless networks in a plethora of propagation environments.
The proposed framework has the potential of  simplifying the cumbersome
procedure  of analyzing the coverage probability  and allowing the   unification of single- and multi-antenna networks  through  compact analytical representations.  By harnessing this key feature,  we develop a novel statistical machinery  to study the  scaling laws  of wireless networks densification  considering general channel power distributions including small-scale fading and shadowing as well as associated beamforming and array gains due to the use of multiple antenna. We further formulate the
relationship between network density,  antenna height, antenna array seize and carrier frequency showing how the coverage probability can be
maintained with ultra-densification.
From a system design perspective, we show that, if multiple antenna base stations are deployed at higher frequencies, monotonically increasing the coverage probability by means of ultra-densification is possible, and this without lowering the antenna height.
 Simulation results substantiate  performance trends leveraging network densification and  antenna deployment and configuration against path loss models and signal-to-noise plus interference  thresholds.

\let\thefootnote\relax\footnotetext{Work supported by the Discovery Grants and the CREATE PERSWADE (www.create-perswade.ca) programs of NSERC and a Discovery Accelerator Supplement (DAS) Award from NSERC. Part of this work has been published in the IEEE WCNC 2020 \cite{trigui11}.
}
\end{abstract}


\section{Introduction}
Chiefly urged by the unfolding mobile data deluge, a radical design make-over of cellular systems enabled by the so-called  network densification and heterogeneity, primarily through the provisioning of small cells,  has become an  extremely active and promising research topic \cite{and1}-\!\!\cite{trigui}. While small-cell densification has been recognized as a promising
solution to boost capacity and enhance coverage with low cost and power-efficient infrastructure in 5G networks,  it also paves the way for reliable and high capacity millimeter wave (mmWave) communication
and directional beamforming \cite{and1}. Nevertheless, there has been noticeable divergence between the above outlook and conclusions of various studies on the fundamental limits of network densification, according to which  the latter may eventually stop, at a certain point, delivering
significant capacity gains \cite{and}-\!\!\cite{lee}.

In this respect, several valuable contributions leverage
stochastic geometry (SG) to investigate ultra-dense networks performance
under various pathloss  and propagation models \cite{and}-\cite{renzo}. In the single-input single-output (SISO) context, conflicting findings
based on various choices of path-loss models have identified that the signal-to-noise plus interference (SINR) invariance property, which enables a potentially infinite aggregated data rate resulting from network densification based on the power-law model  \cite{and}-\!\!\cite{trigui1}, vanishes
once a more physically feasible path loss model is considered. In the latter case, \cite{dens}-\!\!\cite{ding} showed that the coverage probability attains a
maximum point before starting to decay when the network
becomes denser.
Most recently, the authors of \cite{ding}, \cite{atzeni} and \cite{filo}  have investigated the limits of network densification
when the path-loss model includes the antenna height. Besides  invalidating the SINR invariance property in this case, these works find that  by lowering the antenna
height the coverage drop  due to ultra densification can be totally offset, thereby improving  the network capacity.\\ Motivated by the tractability of the considered system models, most of the previous works assumed the scenario of  exponential-based distributions for the channel gains (e.g., integer
fading parameter-based power series \cite{renzo1},\cite{gupta}, \cite{sir}, and Laguerre
polynomial series in \cite{d2dsku}) and unbounded power law models, while the few noteworthy studies that incorporate general fading, shadowing and path-loss models often lead to complex mathematical frameworks that fail to explicitly unveil the relationship
between network density and system performance \cite{renzo}, \cite{trigui}, \cite{trigui1}. Moreover, although some works investigated
the effect of pathloss singularity \cite{sawy},\!\cite{het2},\!\cite{het3} or boundedness \cite{and3},\!\cite{dilhon}, the incorporation of the combined effect of path-loss  and generalized fading channel models is usually ignored. This has  entailed divergent or even contrasting conclusions  on the fundamental limits of network densification \cite{and3},\!\cite{ding},\!\cite{atzeni}. More importantly, additional work is necessary to investigate advanced communication and signal processing techniques, e.g., massive multiple-input-multiple-output (MIMO),  coordinated multipoint (CoMP) and mmWave communications \cite{vm1}-\!\!\cite{vm4}  that are expected to enhance the channel gain.

Motivated by the above background, our work proposes a unified and comprehensive multiple-parameter Fox's H fading model for general multi-path and/or shadowing distributions, which is proved to enable a
tractable analysis of dense networks. The main objective of this paper, in particular, is to introduce a  non model-specific channel model that leads to a new unified approach  to asses the performance of dense networks. To this end,  the proposed framework is based on Fox's H transform theory and the Mellin-Barnes integrals along with SG to investigate the performance limits of network densification under realistic pathloss models and general channel power distributions, including propagation impediments and transmission gains due to the antenna pattern and beamforming, which are particulary relevant in multi-antenna settings.  Several works \cite{lee}-\!\!\cite{xu} studied different
performance metrics to characterize the performance of multi-antenna cellular networks, yet under the assumption of the standard
power-law path-loss model, since it leads to tractable analysis. In this paper, by leveraging a novel methodology of analysis that is compatible with a wide class of path-loss models, including the antenna height, we are able to study the achievable performance
 of multi-antenna networks  and understand how scaling the deployment density of the base stations (BSs) helps maintain the per user-coverage  in  dense networks.  The main contributions of this paper are the following:

\begin{itemize}
  \item
  We introduce a unified analytical framework for analyzing heterogeneous  networks under general Fox's H distributed channel models and both unbounded and  bounded path-loss models.  Closed-form expressions for the coverage probability and corresponding scaling laws allow us to confirm that the path-loss model plays a significant role in determining the network performance, a result corroborated by other recent works on ultra-dense networks \cite{and}-\!\!\cite{xu}.

  \item By exploiting the proposed Fox's H-based channel power representation, we obtain an analytical framework for the coverage probability  in  multi-antenna networks  that is shown to preserve the tractability of the single-antenna case.  The asymptotic performance limits of multi-antenna networks are derived in closed-form showing that there is potential for improving the
scaling laws of the coverage by increasing the number of  BS antennas.

  \item Harnessing  the tractability of the developed analytical model,  the impact of network densification   is investigated by considering advanced transmission techniques, such as MIMO and directional beamforming, and by considering the effect of high transmission frequencies (e.g., mmWave). We show that maintaining  the maximum coverage  is possible by deploying multiple antenna at the BS and by operating at higher frequency bands, and without lowering the BS height. The obtained scaling laws  provide valuable system design guidelines for optimizing general  networks deployment.

\end{itemize}

The rest of the paper is organized as follows. In Section II, we present the system model and the modeling assumptions. In Section III, we introduce our approach to obtain exact closed-form expressions and scaling laws for the coverage probability. Section IV is  focused
on multi-antenna BSs and single-antenna users, i.e., MISO networks. Applications of the obtained  coverage expressions in different wireless communication scenarios are detailed in order to leverage the full potential of network densification. Numerical and simulation results are illustrated  in Section V. Finally, Section IV concludes the paper.

 \section{System Model}
 We consider the downlink transmission of  a ${\cal T}$-tier heterogeneous
wireless network. We focus on the performance analysis of
a typical user equipment (UE) which is assumed, without loss of generality, to be located at the origin  and to be served by the $k$-th tier. Hence, its  SINR is given by
\begin{equation}
\text{SINR}_{k}=\frac{ L(r_{k})g_{x_k}}{\sum_{j=1}^{{\cal T}}\sum_{r_i\in \Phi_j \setminus r_k}\widetilde{P}_i L(r_i)g_{x_i}+\sigma^{2}_k},
\label{eq1}
\end{equation}
where the following notation is used:

\begin{itemize}
  \item $L(r_{i})$ is the large-scale channel gain between the  typical UE  and  the BS at distance $r_i$,
where  $L(r)=r^{-\alpha}$ for an unbounded path-loss and $L(r)=(1+r)^{-\alpha}$ for a bounded one.
\item $\tilde P_i = \frac {P_i} {P_{k}}$ is the power of the $i$-th BS normalized by the power of the BS with index $k$ serving the typical UE.
  \item $\sigma^{2}_k$ is   the normalized noise power defined as $\sigma^{2}_k=\frac{\sigma^{2}}{P_k}$
  \item $g_{x_k}$ is the channel power gain for the desired signal from
the associated transmitter  located at $x_{k}$.
Different channel distributions and MIMO techniques
lead to different distributions for $g_{x_k}$ . In this paper, a general type of distribution is assumed for $g_{x_k}$, as in Assumption 1.

\textit{Assumption 1:}  The channel power gain $g_{x_k}$  for the typical UE has a Fox's H  distribution, i.e., $g_{x_k}\thicksim H_{p,q}^{u,v}(x;{\cal P}_k)$,
with  the  parameter sequence ${\mathcal P}_k= (\kappa_k, c_k, a_k, b_k, A_k, B_k)$ and probability density function (pdf) \cite{mathai}
\begin{equation}
f_{g_{x_k}}(x)=  \kappa_k H_{p,q}^{u,v}\left[ c_k x \left|
\begin{array}{ccc} (a_k, A_k)_p \\ (b_k,B_k)_q \end{array}\right. \right], \quad x\geq0.
\label{eqh}
\end{equation}
The main  advantage of the H-function
representation for statistical distributions is that any algebraic combination involving
products, quotients, or powers of any number of independent
positive continuous random variables can be written
as an H-function distribution. Indeed,  the Fox's H function distribution   captures composite effects of multipath
fading and shadowing, subsuming a wide variety of important or generalized fading distributions
 adopted in wireless communications such as $\alpha$-$\mu$\footnote{The $\alpha$-$\mu$ distributions can be attributed to exponential, one-sided
Gaussian, Rayleigh, Nakagami-m, Weibull and Gamma fading distributions
by assigning specific values for $\alpha$ and $\mu$.},  $N$-Nakagami-$m$, (generalized) ${\cal K}$-fading, and Weibull/gamma fading, and  the Fisher-Snedecor F-S $\cal F$ (\cite{cotton} and  \cite{aloui} and references therein).
\item $\alpha$ is the path loss exponent.
\item $g_{x_i}$ is the interferer's power gain from the interfering transmitter located at $x_i$.  In the proposed framework, we assume that $g_{x_i}$,  $i \in \{1,\ldots {\cal T}\}$
 are non-negative random variables that are independent and
identically distributed according to (\ref{eqh}).
 \end{itemize}

\section{  UNIFIED ANALYTICAL FRAMEWORK}
In this section, we derive the complementary cumulative distribution function (ccdf) of the SINR, also
called the coverage probability, in single-antenna networks. The obtained framework is utilized to analyze multi-antenna networks in general
settings in the next section.
\subsection{Coverage Analysis in Closest-BS-Association-Based Cellular Networks}

\textit{Proposition 1:}
When the locations of the BSs are modeled as a Poisson point process (PPP) \cite{ren3} and the nearest-BS association is adopted, the SINR coverage probability at the typical UE for an unbounded path-loss model and the SINR thresholds $\beta_k$, $k\in\{1,\ldots, {\cal T}\}$, is given by
\begin{eqnarray}
\!\!\!\!\!\!\!{\cal C}^{\cal U}\!\!\!&=& \!\!\!\pi \delta \sum^{{\cal T}}_{k=1} \lambda_k \left(\frac{P_k}{\sigma_k^{2}}\right)^{\delta}\int_{0}^{\infty}\frac{1}{\xi^{2+\delta}}H_{q,p+1}^{v,u}\left(\xi, {\cal P}^{k}_{\cal U}\right)\nonumber\\ &&  \!\!\!\!\!\!\!\!\!\!\!\!\!\!\!\!\!\!\!\!\!\!\!\!{\mathcal  H}_{1,1}^{1,1}\!\!\left(\!\!\left(\frac{P_k}{\xi\sigma^{2}_k}\right)^{\delta}\!\!\sum_{j=1}^{{\cal T}}\!\!\! \pi \lambda_j \widetilde{P}^{\delta}_j \!\left(1\!+\!\delta \xi H_{q+2,p+3}^{v+1,u+2}\!\!\left(\xi, {\cal P}^{{\cal I}_j}_{\cal U}\right)\!\right)\!,\!{\cal P}_{\delta}\!\!\right)\!d\xi,
\label{cb}
\end{eqnarray}
where $\delta=\frac{2}{\alpha}$, ${\cal P}_{\delta}=\left(1,1, 1-\delta, 0, \delta, 1\right)$, with
\begin{equation}
 {\cal P}^{k}_{\cal U}=\!\!\left(\kappa_k {\cal \beta}_k, \frac{1}{c_k {\cal \beta}_k}, 1\!-\!b_k, (1\!-\!a_k, 1), B_k, (A_k,1)\right),
 \end{equation}
 and
 \begin{eqnarray}
{\cal P}^{{\cal I}_j}_{\cal U}\!\!\!\!\!&=&\!\!\!\!\!\!\bigg(\frac{\kappa_j}{c_j^{2}}, \frac{1}{c_j}, (1\!-\!b_j\!-\!2B_j,0,\delta),(0, 1\!-\!a_j\!-\!2A_j, -1,\delta\!-\!1),\nonumber \\ && (B_j,1,1), (1,A_j,1,1)\bigg).
\end{eqnarray}

\textit{Proof:} See Appendix A.

The main assumptions in Proposition 1 are the Fox's H distributed signal and interference channel power gains and the standard power-law unbounded path loss model. The unbounded power-law path loss is known to be inaccurate for short distances, due to the singularity
at the origin, which affects the scaling laws of the coverage probability \cite{sawy}. Next,  a more physically feasible path-loss model is considered.

\textit{Proposition 2:} When  a bounded path-loss model is adopted, the coverage probability of cellular networks  based on the nearest-BS association strategy is given in (6)-(8),  that are shown at the top of this page.
%
\begin{figure*}
\begin{eqnarray}
{\cal C}^{\cal B}&=&\sum_{k=1}^{{\cal T}}\lambda_k \int_{0}^{\infty}e^{-\sum_{j\in {\cal T}} \pi \lambda_j \widetilde{P}^{\delta}_j\delta \xi (\Psi_1-\Psi_2)}\frac{H_{q,p+1}^{v,u}\!\!\left(\xi, {\cal P}^{k}_{\cal B}\right)}{\xi^{2}\sum_{j=1}^{{\cal T}} \pi \lambda_j \widetilde{P}^{\delta}_j\delta \xi (\Psi_1+\Psi_2)}H_{1,1}^{1,1}\left(\!\!\frac{\sum_{j=1}^{{\cal T}} \lambda_j \widetilde{P}^{\delta}_j\left(1\!+\!\delta \xi \Psi_1\right)}{\sum_{j=1}^{ {\cal T}} \lambda_j \widetilde{P}^{\delta}_j\delta \xi(2 \Psi_1\!+\!\Psi_2)}, {\cal \widetilde{P}}_{\delta}\!\!\right)\!d\xi,
\label{mgf1}
\end{eqnarray}
\text{where} ${\cal \widetilde{P}}_{\delta}=\left(1,1, -1, 0, 2, 1\right)$, ${\cal P}^{k}_{\cal B}={\cal P}^{k}_{\cal U}$ \text{and} $\Psi_x=H_{q+2,p+3}^{v+1,u+2}\left(\xi, {\cal P}^{x,{\cal I}}_{\cal B}\right), x\in\{1,2\}$ \text{with}
\begin{equation} {\cal P}^{1,j,{\cal I}}_{\cal B}= \bigg(\frac{\kappa_j}{c_j^{2}}, \frac{1}{c_j}, (1-b_j-2B_j,0,\delta),(0, 1-a_j-2A_j, -1,\delta-1), (B_j,1,1), (1,A_j,1,1)\bigg)
\end{equation}
\text{and}
\begin{equation}{\cal P}^{2,j,{\cal I}}_{\cal B}=\bigg(\frac{\kappa_j}{c_j^{2}}, \frac{1}{c_j}, \left(1\!-\!b_j\!-\!2B_j,0,\frac{\delta}{2}\right),\left(0, 1\!-\!a_j\!-\!2A_j, -1,\frac{\delta}{2}\!-\!1\right), (B_j,1,1), (1,A_j,1,1)\bigg) \end{equation}
\hrulefill
\end{figure*}

\textit{Proof:} See Appendix B.

\textit{Remark 1:} For arbitrary distributions for the channel gain, the coverage expressions in (\ref{cb}) and (\ref{mgf1}) are independent of the $n$-th derivative of the Laplace transform of the aggregate interference, $n\in[0,\infty)$, while accurately reflecting the behavior
of multi-tiers networks in all operating regimes without the need of applying approximations or upper bounds.
 Compared with the coverage approximations in \cite{d2dsku}, \cite{gupta},  and \cite{sir}  and expressions in \cite{renzo},\!\cite{renzo1}, the proposed approach yields a
more compact analytical result for the coverage probability, where only an
integration of Fox's-H functions is needed thanks to the novel  handling of fading distributions. Table II lists some commonly-used channel fading distributions and
the corresponding expression for ${\cal C}$.\\ It is worthy to note that  the proposed framework  can be extended to other network models, for example, where the
transmitters are spatially distributed according to other point
processes \cite{ren4}, \cite{ren5},  notably including non-Poisson models \cite{het3}, or under multi-slope path loss models \cite{kon}, \cite{het2}. Hence, the results of this paper allow  an exact and tractable approximation for the coverage probability of any stationary
and ergodic point process \cite{het3}, \cite{ren4}. In the next section, we show the usefulness of the proposed approach for obtaining  insightful design guidelines for multi-antenna and  mmWave networks.


\begin{table*}
\caption{COVERAGE  PROBABILITY OF SOME
WELL-KNOWN FADING CHANNEL MODELS BASED ON THE CLOSEST-BS STRATEGY}
\centering
\begin{tabular}{p{2.5in} p{4.53in}}
  \hline\hline
  \textbf{Instantaneous Fading Distribution} & \textbf{Coverage Probability} ${\cal C}^{\cal U}$ \\ \hline\hline
\textbf{Gamma Fading}

 $f_g(z)=\frac{m}{\Gamma(m)} H_{0,1}^{1,0}\left[ m z \left|
\begin{array}{ccc} - \\ (m-1,1) \end{array}\right. \right]$
 & $\begin{aligned} &{\cal C}^{\cal U}=\pi \delta \sum^{{\cal T}}_{k=1} \frac{\lambda_k m_k \beta_K}{\Gamma(m_k)} \left(\frac{P_k}{\sigma_k^{2}}\right)^{\delta}\int_{0}^{\infty}\frac{H_{1,1}^{0,1}\left[ \frac{\xi}{m_k \beta_k} \left|
\begin{array}{ccc} (2-m_k,1) \\ (1,1) \end{array}\right. \right]}{\xi^{2+\delta}} \\ & {\mathcal  H}_{1,1}^{1,1}\left(\!\!\frac{\left(\frac{P_k}{\sigma^{2}_k}\right)^{\delta}}{\xi^{\delta}}\!\!\sum_{j=1}^{\cal T}\!\!\! \pi \lambda_j \widetilde{P}^{\delta}_j \!\left(1\!+\!\frac{\delta \xi}{m_j\Gamma(m_j)} H_{3,3}^{1,3}\!\!\left[ \frac{\xi}{m_j} \left|
\begin{array}{ccc} (-m_j,1), (0,1), (\delta,1) \\ (0,1), (-1,1),(\delta-1,1) \end{array}\right. \right]\right)\!,\!{\cal P}_{\delta}\!\!\right)\!d\xi\end{aligned}$ \\\\
  \hline
\textbf{Generalized Gamma}

$\begin{aligned} & f_g(z)=\frac{\mu}{\Gamma(m)} H_{0,1}^{1,0}\left[ \mu z \left|
\begin{array}{ccc} - \\ (m-\frac{1}{\eta},\frac{1}{\eta}) \end{array}\right. \right]\end{aligned}$
{\small where $\mu=\frac{\Gamma(m+\frac{1}{\eta})}{\Gamma(m)}$.}  &$\begin{aligned} &{\cal C}^{\cal U}=\pi \delta \sum^{{\cal T}}_{k=1} \frac{\lambda_k \mu_k \beta_K}{\Gamma(m_k)} \left(\frac{P_k}{\sigma_k^{2}}\right)^{\delta}\int_{0}^{\infty}\frac{H_{1,1}^{0,1}\left[ \frac{\xi}{\mu_k \beta_k} \left|\begin{array}{ccc} (1+\frac{1}{\eta_k}-m_k,\frac{1}{\eta_k}) \\ (1,1) \end{array}\right. \right]}{\xi^{2+\delta}}\\ & {\mathcal  H}_{1,1}^{1,1}\left(\!\!\frac{\left(\frac{P_k}{\sigma^{2}_k}\right)^{\delta}}{\xi^{\delta}}\!\!\sum_{j=1}^{\cal T}\!\!\! \pi \lambda_j \widetilde{P}^{\delta}_j \!\left(1\!+\!\frac{\delta \xi}{\mu_j\Gamma(m_j)} H_{3,3}^{1,3}\!\!\left[ \frac{\xi}{\mu_j} \left|\begin{array}{ccc} (1-\frac{1}{\eta_j}-m_j,1), (0,1), (\delta,1) \\ (0,1), (-1,1),(\delta-1,1) \end{array}\right. \right]\right)\!,\!{\cal P}_{\delta}\!\!\right)\!d\xi\end{aligned}$\\\\ \hline
  \textbf{Power of Nakagami-$n$ (Rice)}

  $\begin{aligned} & f_g(z)=\sum_{k=0}^{\infty} \frac{\Psi_k m_k}{\Gamma(m_k)\Omega_k} H_{0,1}^{1,0}\left[ \frac{m_k}{\Omega_k} z \left|
\begin{array}{ccc} - \\ (m_k-1,1) \end{array}\right. \right]\end{aligned}$   {\small with $m_k = k + 1$ and $\Omega_k= \frac{k + 1}{1+K_R}$, where $K_R$ is the Rician factor.}   & $\begin{aligned} &{\cal C}^{\cal U}=\pi \delta \sum^{{\cal T}}_{k=1}\underset{K_k\longrightarrow\infty}{\lim}\sum_{t=0}^{K_k} \frac{\Psi_t m_t}{\Gamma(m_t)\Omega_t \lambda_k \beta_k}  \left(\frac{P_k}{\sigma_k^{2}}\right)^{\delta}\int_{0}^{\infty}\frac{H_{1,1}^{0,1}\left[ \frac{\xi \Omega_t}{m_t \beta_k} \left|
\begin{array}{ccc} (2-m_t,1) \\ (1,1) \end{array}\right. \right]}{\xi^{2+\delta}} \\ & {\mathcal  H}_{1,1}^{1,1}\!\!\left(\!\!\frac{\left(\frac{P_k}{\sigma^{2}_k}\right)^{\delta}}{\xi^{\delta}}\!\!\sum_{j=1}^{{\cal T}}\!\!\! \pi \lambda_j \widetilde{P}^{\delta}_j \!\left(1\!+\!\sum_{t=0}^{\infty} \frac{\Psi_t \Omega_t \delta \xi}{\Gamma(m_t)m_t} H_{3,3}^{1,3}\!\!\left[ \frac{\xi \Omega_t}{m_t} \left|
\begin{array}{ccc} (-m_t,1), (0,1), (\delta,1) \\ (0,1), (-1,1),(\delta-1,1) \end{array}\right. \right]\!\right)\!,\!{\cal P}_{\delta}\!\!\right)\!d\xi\end{aligned}$,    {\small where $\Psi_k=K_R^{k}e^{-K_R}/\Gamma(k+1)$. }\\\\
\hline
\textbf{Lognormal Fading}

   $\begin{aligned} & f_g(z)=\sum_{n=0}^{N} \frac{w_n}{\omega_n} H_{0,0}^{0,0}\left[ \frac{z}{\omega_n}\left|
\begin{array}{ccc} - \\- \end{array}\right. \right]\end{aligned}$ {\quad\quad\quad\quad \small where $\omega_n=10^{\sqrt{2}\sigma u_n+\mu}$,  while $u_n$ and $w_n$ represrent the weight factors and the zeros of the $N$-order Hermite polynomial [6, Table 25.10]}. & $\begin{aligned} &{\cal C}^{\cal U}=\pi \delta \sum^{{\cal T}}_{k=1}\sum_{t=0}^{N_k} \frac{w_t }{\omega_t \lambda_k \beta_k}  \left(\frac{P_k}{\sigma_k^{2}}\right)^{\delta}\int_{0}^{\infty}\frac{H_{0,1}^{0,0}\left[ \frac{\xi \omega_t}{ \beta_k} \left|
\begin{array}{ccc} - \\ (1,1) \end{array}\right. \right]}{\xi^{2+\delta}}\\ & {\mathcal  H}_{1,1}^{1,1}\!\!\left(\!\!\frac{\left(\frac{P_k}{\sigma^{2}_k}\right)^{\delta}}{\xi^{\delta}}\!\!\sum_{j=1}^{{\cal T}}\!\!\! \pi \lambda_j \widetilde{P}^{\delta}_j \!\left(1\!+\!\sum_{t=0}^{N_j} \frac{ w_t \delta \xi}{\omega^{2}_t} H_{2,3}^{1,2}\!\!\left[ \xi \omega_t \left|
\begin{array}{ccc}  (0,1), (\delta,1) \\ (0,1), (-1,1),(\delta-1,1) \end{array}\right. \right]\!\right)\!,\!{\cal P}_{\delta}\!\!\right)\!d\xi\end{aligned}$ \\\\
\hline
\textbf{Fisher-Snedecor Fading }

  $\begin{aligned} & f_g(z)=\frac{m}{m_s \Gamma(m_s)\Gamma(m)} H_{1,1}^{1,1}\left[ \frac{m z}{m_s} \left|
\begin{array}{ccc} (-m_s,1) \\ (m-1,1) \end{array}\right. \right]\end{aligned}$
 & $\begin{aligned} &{\cal C}^{\cal U}=\pi \delta \sum^{{\cal T}}_{k=1} \frac{\lambda_k m_k \beta_K}{m_{s_k}\Gamma(m_k)\Gamma(m_{s_k})} \left(\frac{P_k}{\sigma_k^{2}}\right)^{\delta}\int_{0}^{\infty}\frac{H_{1,2}^{1,1}\left[ \frac{\xi m_{s_k}}{m_k \beta_k} \left|
\begin{array}{ccc} (2-m_k,1) \\ (1+m_{s_k},1), (1,1) \end{array}\right. \right]}{\xi^{2+\delta}}\\ & {\mathcal  H}_{1,1}^{1,1}\left(\!\!\frac{\left(\frac{P_k}{\sigma^{2}_k}\right)^{\delta}}{\xi^{\delta}}\!\!\sum_{j=1}^{{\cal T}}\!\!\! \pi \lambda_j \widetilde{P}^{\delta}_j \!\left(1\!+\!\frac{\delta \xi}{m_j\Gamma(m_j)} H_{3,3}^{1,3}\!\!\left[ \frac{\xi}{m_j} \left|
\begin{array}{ccc} (-m_j,1), (0,1), (\delta,1) \\ (0,1), (-1,1),(\delta-1,1) \end{array}\right. \right]\right)\!,\!{\cal P}_{\delta}\!\!\right)\!d\xi\end{aligned}$ \\\\
\hline
\end{tabular}
\end{table*}


%

\subsection{Coverage  Analysis in Strongest-BS-Association-Based Cellular Networks}
The strongest-BS association rule, according to which   the serving  BS  is  the  one  that  provides  the  maximum  signal-to-interference (SIR)\footnote{\cite{dilhon} showed that self-interference dominates noise in
 typical heterogeneous networks under strongest-BS association. Therefore, we ignore noise in the rest of this
section.},   can  be  particularly  advantageous for application to scenarios in which    the  closest-BS association strategy  may provide poor performance due  to  severe blockage.  Also, the strongest-BS association criterion may yield performance bounds for other, more practical, cell association strategies.

\textit{Proposition 3:}
When  the strongest-BS association is adopted, the SIR coverage probability of the typical UE, given the SIR thresholds $\beta_k$, $k\in\{1,\ldots, {\cal T}\}$, is given by
\begin{eqnarray}
\!{\cal C}&=& 2\pi \sum_{k=1}^{{\cal T}} \frac{\kappa_k \lambda_k}{c_k} \int_{0}^{\infty}r_k \Upsilon(r_k)d_{r_k},\nonumber \\
&=& \frac{\pi}{C(\delta)}  \sum_{k=1}^{{\cal T}}\frac{\lambda_k \beta_k^{-\delta}\Lambda_k }{\sum_{j\in {\cal T}}\lambda_j \widetilde{P}_j^{\delta} \Lambda_j},
\label{cuas}
\end{eqnarray}
with
\begin{eqnarray}
\!\Upsilon(r_k)\!&=&\!\nonumber\\&& \!\!\!\!\!\!\!\!\!\!\!\!\!\!\!\!\!\!\!\!\!\!\!\!\!\!\!\!\!\!\!\!\!\!\!\!H_{p+1,q+1}^{u+1,v}\!\!\left[\sum_{j=1}^{{\cal T}}\!\! \frac{\pi r_k^{2} \lambda_j \Gamma(1\!-\!\delta)\Lambda_j }{\widetilde{P}^{-\delta}_j ({c_k} \beta_k)^{-\delta}}\!\left|\!\!
\begin{array}{ccc}\! (a_k\!\!+\!\!A_k, \delta A_k), (1, \delta) \\ (0,1), (b\!+\!B_k, \delta B_k) \end{array}\right.\!\!\! \!\!\right],
\label{eqY}
\end{eqnarray}
where
$C(\delta)=\pi^{2}\delta~{\rm csc}(\pi \delta)$ and
\begin{eqnarray}
\Lambda_j&=&\frac{\kappa_j}{c_j^{\delta+1}}\frac{\prod_{t=1}^{u}\Gamma\left(b_{j_{t}}+(1+\delta)B_{j_{t}}\right)
}{\prod_{t=u+1}^{p}\Gamma\left(1-b_{j_{t}}-(1+\delta)B_{j_{t}}\right)}\nonumber\\ && \times \frac{\prod_{k=1}^{v}\Gamma\left(1-a_{j_{k}}-(1+\delta)A_{j_{k}}\right)}{\prod_{k=v+1}^{p}\Gamma\left(a_{j_{k}}+(1+\delta)A_{j_{k}}\right)}.
\label{lambdaj}
 \end{eqnarray}

\textit{Proof:} The proof follows from Appendix C along with the fact that
\begin{eqnarray}
{\mathcal E}_{r_k} \left[\Upsilon(r_k)\right]\!\!\!&=&\!\!\!2 \pi \lambda_k \int_{0}^{\infty}\!\! \!r_k H_{p+1,q+1}^{u+1,v}\Bigg[\!\sum_{j=1}^{{\cal T}}\!\! \frac{\pi r_k^{2} \lambda_j \Gamma(1\!-\!\delta)\Lambda_j }{\widetilde{P}^{-\delta}_j\left({c_k} \beta_k\right)^{-\delta}}\!\nonumber \\ && \left|\!\!
\begin{array}{ccc}\! (a_k\!\!+\!\!A_k, \delta A_k), (1, \delta) \\ (0,1), (b\!+\!B_k, \delta B_k) \end{array}\right.\!\!\! \!\!\Bigg]d_{r_k}.
\end{eqnarray}
Then, applying the transformation $H_{p,q}^{m,n}\big[x \big|
\begin{array}{ccc} (a_i, k A_j)_p \\ (b_i, k B_j)_q \end{array}\big. \big]=\frac{1}{k}H_{p,q}^{m,n}\big[x^{\frac{1}{k}} \big|
\begin{array}{ccc} (a_i, A_j)_p \\ (b_i, B_j)_q \end{array}\big. \big]$, $k>0$ and the Mellin transform in \cite{mathai}, we obatin
\begin{eqnarray}
{\mathcal E}_{r_k} \left[\Upsilon(r_k)\right]&=&  \frac{\Gamma(1-\delta)^{-1}}{\Gamma(1+\delta)} \frac{c_k \beta_k^{-\delta}\frac{\Lambda_k}{\kappa_k} }{\sum_{j\in {\cal T}}\lambda_j \widetilde{P}_j^{\delta} \Lambda_j}.
\label{eqe1}
\end{eqnarray}
Finally, plugging (\ref{eqe1}) into (\ref{outcu}) yields the desired result after some manipulations.

\textit{Remark 2:}
As shown in (\ref{outcu}), the main task in deriving the coverage
probability in cellular networks under the strongest-BS cell association criterion is to calculate $\Lambda$. In Table II, we show the coverage probability for the strongest-BS association criterion when various special cases of the Fox's H-function distribution are considered. Notably, (\ref{cuas}) is instrumental in evaluating the impact  of the number of tiers or their relative densities,  transmit powers, and  target SIR  over generalized fading scenarios. This result complements existing valuable coverage studies of cellular networks over generalized fading \cite[Proposition 1]{sir}, \cite[Corollary 1]{dilhon}.

\subsection{Coverage Analysis in Ad Hoc Networks}
Ad hoc networks with short range transmission are, from an architecture perspective, similar to device-to-device (D2D) communication networks where Internet of Things (IoT) devices communicate directly over the regular cellular spectrum but without using the BSs.
In ad hoc networks, the communication distance $r_k$ between the typical receiver and its associated transmitter in the $k$-th tier is
assumed to be fixed and  independent of the set of interfering transmitters and their densities.

\textit{Proposition 4:} The coverage probability of ad hoc networks over the Fox's H fading channel is given by
\begin{eqnarray}
\!{\cal C}&=&\sum_{k=1}^{{\cal T}} \frac{\kappa_k}{c_k} \Upsilon(r_k).
\label{outcu}
\end{eqnarray}
where $\Upsilon(r_k)$ is given in (\ref{eqY}).

We note  that the coverage probability in ad hoc networks involves finite summation of Fox's H functions which can be efficiently evaluated \cite{trigui1}.   Overall, the obtained analytical expressions are easier to compute  than
existing results \cite{and}-\!\!\cite{renzo1}, \cite{ding}-\!\!\cite{lee} that contain multiple nested integrals.

\begin{table*}
\caption{ COVERAGE  PROBABILITY OF SOME
WELL-KNOWN FADING CHANNEL MODELS BASED ON THE STRONGEST-BS ASSOCIATION}
\centering
\begin{tabular}{p{2.3in} p{3.5in}}
  \hline
  \textbf{Instantaneous Fading Distribution} & \textbf{Coverage Probability} ${\cal C}^{\cal U}$ \\ \hline \\
\textbf{Gamma Fading}
  & $\begin{aligned} &{\cal C}^{{\cal U}}&=&\frac{\pi}{C(\delta)} \sum_{k=1}^{{\cal T}}\frac{\lambda_k \beta_k^{-\delta}\frac{\Gamma\left(m_k+\delta\right)}{\Gamma\left(m_k\right) m_k^{\delta}}  }{\sum_{j=1}^{{\cal T}}\lambda_j \widetilde{P}_j^{\delta} \frac{\Gamma\left(m_j+\delta\right)}{\Gamma\left(m_j\right) m_j^{\delta}}}.\end{aligned}$ \\ \\
  \hline \\
\textbf{Generalized Gamma} & $\begin{aligned} &{\cal C}^{{\cal U}}&=&\frac{\pi}{C(\delta)} \sum_{k=1}^{{\cal T}}\frac{\lambda_k \beta_k^{-\delta}\frac{\Gamma(\mu_k)^{\delta-1}}{\Gamma\left(\mu_k+\frac{1}{\alpha_k}\right)^{\delta}} \Gamma\left(\mu_k+\frac{\delta}{\alpha_k}\right) }{\sum_{j=1}^{{\cal T}}\lambda_j \widetilde{P}_j^{\delta} \frac{\Gamma(\mu_j)^{\delta-1}}{\Gamma\left(\mu_j+\frac{1}{\alpha_j}\right)^{\delta}} \Gamma\left(\mu_j+\frac{\delta}{\alpha_j}\right)}.\end{aligned}$\\ \\ \hline\\
  \textbf{Power of Nakagami-$n$ (Rice)}
  & $\begin{aligned} &{\cal C}^{{\cal U}}&=&\frac{\pi}{C(\delta)} \sum_{k=1}^{\cal T}\frac{\lambda_k \beta_k^{-\delta}e^{-K_{R_k}}\sum_{t=0}^{\infty}\frac{K_{R_k}^{t}\Gamma(m_t+\delta)}{\Gamma(t+1)\Gamma(m_t)}\left(\frac{\Omega_t}{m_t}\right)^{\delta} }{\sum_{j=1}^{\cal T}\lambda_j e^{-K_{R_j}} \widetilde{P}_j^{\delta} \sum_{t=0}^{\infty}\frac{K_{R_j}^{t}\Gamma(m_t+\delta)}{\Gamma(t+1)\Gamma(m_t)}\left(\frac{\Omega_t}{m_t}\right)^{\delta} }.\end{aligned}$ \\ \\
\hline\\
\textbf{  Lognormal Fading}  & $\begin{aligned} &{\cal C}^{{\cal U}}&=&\frac{\pi}{C(\delta)} \sum_{k=1}^{\cal T}\frac{\lambda_k \beta_k^{-\delta} \sum_{n=0}^{N_k}w_n 10^{\delta(\sqrt{2}\sigma_k u_n+\mu_k)} }{\sum_{j=1}^{\cal T}\lambda_j \widetilde{P}_j^{\delta}\sum_{n=0}^{N_j}w_n 10^{\delta(\sqrt{2}\sigma_j u_n+\mu_j)}}.\end{aligned}$  \\ \\
\hline\\
\textbf{ Fisher-Snedecor Fading}   & $\begin{aligned} &{\cal C}^{{\cal U}}&=&\frac{\pi}{C(\delta)} \sum_{k=1}^{\cal T}\frac{\lambda_k \beta_k^{-\delta}\frac{m_{s_k}^{\delta}\Gamma\left(m_k+\delta\right)\Gamma(m_{s_k}-\delta)}{\Gamma(m_{s_k})\Gamma\left(m_k\right) m_k^{\delta}}  }{\sum_{j=1}^{\cal T}\lambda_j \widetilde{P}_j^{\delta} \frac{m_{s_j}^{\delta}\Gamma\left(m_j+\delta\right)\Gamma(m_{s_j}-\delta)}{\Gamma(m_{s_j})\Gamma\left(m_j\right) m_j^{\delta}} }.\end{aligned}$  \\ \\
\hline
\end{tabular}
\end{table*}


\vspace{-0.2cm}

\subsection{The Impact of Network Densification}
In this section, we exploit the   derived analytical framework  to analyze the coverage scaling laws for  single-antenna  multi-tiers cellular and ad hoc networks. Assuming $\lambda_k=\lambda\rightarrow\infty$, $k=1,\ldots, {\cal T}$, the coverage scaling laws are given in the following text.
\subsubsection{Coverage Scaling Law in Cellular Networks}
The coverage probability of single antenna-cellular networks with an unbounded path-loss model is invariant to the
BS density $\lambda$. Specifically, we have
\begin{eqnarray}
\!\!\!{\cal C}^{\cal U, \infty}\!\!\!\!\!&=& \!\!\!\!\!\sum^{{\cal T}}_{k=1}\!\int_{0}^{\infty}\!\!\!\!\!\!\frac{H_{q,p+1}^{v,u}\left(\xi, {\cal P}^{k}_{\cal U}\right)d\xi}{\xi^{2}\sum_{j=1}^{{\cal T}}\! \widetilde{P}^{\delta}_j\! \left(1\!+\!\delta \xi H_{q+2,p+3}^{v+1,u+2}\!\!\left(\xi, {\cal P}^{{\cal I}}_{\cal U}\right)\right)},
\label{cu1}
\end{eqnarray}
which is obtained by letting $\lambda\rightarrow \infty$ in Proposition 1 and resorting to the asymptotic expansion of the Fox's H function   $H_{1,1}^{1,1}(x;{\cal P}_{\delta})\underset{x\rightarrow\infty}{\approx} \frac{1}{\delta} x^{-1}$ along with  applying \cite[Eq. (1.5.9)]{kilbas}.
We note that (\ref{cu1}) generalizes the SINR invariance property that has been revealed
in some specific settings, e.g., \cite{and3}, \cite{and},  \cite{renzo}, and \cite{dilhon}.\\
Contrary to what the  standard
unbounded path-loss model predicts, the coverage
probability under the bounded path-loss model
scales with $e^{-\lambda}$ and approaches zero with increasing $\lambda$ for general values of $\delta$. This is readily shown in the following asymptotic coverage expression obtained by letting $\lambda\rightarrow \infty$ in Proposition 2, as\footnote{Using the Mellin-Barnes integral representations of $\Psi_1$ and $\Psi_2$ \cite{mathai}, we can easily show that $\Psi_1-\Psi_2>0$.}
\begin{eqnarray}
{\cal C}^{\cal B, \infty}\!\!\!\!&=&\!\!\!\!\!\!\sum_{k=1}^{{\cal T}}\int_{0}^{\infty}e^{-\lambda \sum_{j=1}^{{\cal T}} \pi  \widetilde{P}^{\delta}_j\delta \xi (\Psi_1-\Psi_2)}\nonumber\\ && \!\!\!\!\!\!\!\!\nonumber\\ &&\!\!\!\!\!\!\!\!\!\!\!\!\!\!\!\!\!\!\!\!\!\!\!\!\!\!\times \frac{H_{q,p+1}^{v,u}\left(\xi, {\cal P}^{k}_{\cal B}\right)H_{1,1}^{1,1}\left(\frac{\sum_{j=1}^{{\cal T}} \widetilde{P}^{\delta}_j(1+\delta \xi \Psi_1)}{\sum_{j\in {\cal T}}  \widetilde{P}^{\delta}_j\delta \xi(2 \Psi_1+\Psi_2)}, {\cal P}_{\delta}\right)}{\xi^{2}\sum_{j=1}^{{\cal T}} \pi  \widetilde{P}^{\delta}_j\delta \xi (\Psi_1+\Psi_2)}d\xi.
\label{dens1}
\end{eqnarray}
 Due to the complexity of the bounded  model, its impact was only understood through approximations in \cite{sir} and \cite{dens}  and for fading scenarios  with integer parameters. Thanks to our proposed unified approach,  the impact of ultra densification can be scrutinized in the most comprehensive setting of multi-tier networks under the Fox's $\mathcal{H}$ fading channel.

\subsubsection{Coverage Scaling Law in Ad Hoc Networks}
In ad hoc networks, to the best of our knowledge,  there exists no
works that quantified the effect of densification over generalized fading channels. By exploiting the proposed analytical framework, the coverage scaling law in ad hoc networks is revealed in this paper. First, its is pertinent to  remark that   $g_k \sim \mathcal{H}$-$\{(q,0,p,q), {\mathcal P}\}$ can be assumed in the majority of  fading distributions as shown in Table I. In this case, applying the asymptotic expansion of the Fox's H function \cite[Eq. (1.7.14)]{kilbas} ${\rm H}_{p,q}^{q,0}(x)\sim x^{\frac{\nu+\frac{1}{2}}{\Delta}} \exp\left[-\Delta \left(\frac{x}{\rho}\right)^{1/\Delta}\right]$ to (\ref{outcu}), we obtain
\begin{eqnarray}
\!\!\!{\cal C}\!\!&\underset{\lambda\rightarrow\infty}{\approx}&\! \!\!\sum_{k=1}^{{\cal T}} \frac{\kappa_k}{c_k} \left(\lambda {\cal A} \right)^{\frac{\nu_k+\frac{1}{2}}{\Delta_k}}\exp\left[-\Delta_k \left(\lambda \frac{\!{\cal A}}{\rho_k}\right)^{1/\Delta_k}\right],
\end{eqnarray}
where ${\cal A}=\sum_{j=1}^{{\cal T}}\!\! \frac{\pi r_k^{2}  \Gamma(1\!-\!\delta)\Lambda_j }{\widetilde{P}^{-\delta}_j ({c_k} \beta_k)^{-\delta}}$, $\Delta_k=1+\delta\left(\sum_{j=1}^{q}B_{j_k}-\sum_{j=1}^{p}A_{j_k}-1\right)$, $\rho_k=\delta^{\delta}\prod_{j=1}^{p} (\delta A_{j_k})^{- \delta A_{j_k}}\prod_{j=1}^{q}(\delta B_{j_k})^{-\delta B_{j_k}}$, and $\nu_k=\sum_{j=1}^{q}b_{j_k}-\sum_{j=1}^{p}a_{j_k} +\sum_{j=1}^{q}B_{j_k}-\sum_{j=1}^{p}A_{j_k}+\frac{p-q}{2}-1$ are constants defined in \cite[Eq. (1.1.8)]{kilbas}, \cite[Eq. (1.1.9)]{kilbas}, and \cite[Eq. (1.1.10)]{kilbas}, respectively.\\
In the special case of Gamma fading, i.e.,  $g_{x_k}\sim\texttt{Gamma}(m_k,1)\sim \mathcal{H}$-$\{(1,0,0,1), {\mathcal P}\}$, it can be shown that $\Delta_k=1$, $\rho_k=1$, and $\nu_k=m_k-\frac{3}{2}$, which results in
\begin{eqnarray}
{\cal C}&\underset{\lambda\rightarrow\infty}{\approx}& \!\!\sum_{k=1}^{{\cal T}} \frac{\left(\lambda {\cal A} \right)^{m_k-1}}{\Gamma(m_k)}\exp\left( -\lambda {\cal A} \right).
\label{as1}
\end{eqnarray}
It turns out that  the coverage probability of ad hoc networks in arbitrary Nakagami-$m$ fading (i.e., $g_{x_k}\sim\texttt{Gamma}(m_k,1)$, $k=1,\ldots,{\cal T}$)
is formulated as the  product of an exponential function and a polynomial
function of order ${\cal T}(\underset{k=1,\ldots,\cal{M}}{\max m_k}-1)$ of the transmitter density $\lambda$.
When ${\cal T}=1$, i.e., in single-tier networks, the coverage probability is a product of an exponential function and a power
function of order $m-1$.  In the special case when ${\cal T}=m=1$, i.e.,  in single-tier ad hoc networks over Rayleigh fading channel,  the coverage probability reduces to an exponential function.

\section{ Multi-Antenna  vs. Single-Antenna Networks}
\subsection{Coverage Analysis}
In multi-antenna
networks, the analysis of the coverage probability is more difficult due to more complicated signal and
interference distributions.  However, we emphasize  that, for several MIMO techniques, the associated post-processing signal power gain can include Gamma-type fading \cite{lee}, \cite{vm2}, \cite{renzo2}, \cite{xu} with $g_{x}\sim\texttt{Gamma}(M, \theta)$  where $M$ is typically
related to the number of antennas (e.g. $M=N_t, \theta=1$ for maximum-ratio-transmission (MRT) and $M=N_t, \theta=1/N_t$  for millimeter wave analog beamforming, where $N_t$ is the number of antennas at the transmitter) \cite[Table II]{xu}. Hence, assuming that the signal power is gamma distributed in multi-antenna networks and recognizing that  $f_{g_{x}}(x)=\frac{\theta^{-1}}{\Gamma(M)} H_{0,1}^{1,0}\left[ \frac{x}{\theta} \left|
\begin{array}{ccc} - \\ (M-1,1) \end{array}\right. \right]$, then the Fox's H-based modeling of the coverage presented in Section III can be generalized to muti-anetnna networks analysis.

\subsubsection{Multi-Antenna Cellular Networks}
We consider multiple-input-single-output (MISO) networks using MRT where the BSs in the $k$-th tier are equipped with $N_{t_k}$ antennas. We assume that the channel power gain $g_{x_k}$ for the desired signal is gamma distributed such that $g_{x_k}\sim\texttt{Gamma}(N_{t_k}, 1)$ \cite{xu}. As far as the interference distribution is concerned, we assume  that $g_{x_i}$ are  identically distributed according to an arbitrary  Fox's H   distribution. Hence, Proposition 1  can be generalized  to obtain the coverage probability in multi-antenna cellular networks with arbitrary interference, as
\begin{eqnarray}
{\cal C}\!\!\!&=&\!\!\!\pi \delta \sum^{{\cal T}}_{k=1}\! \frac{ \left(\frac{P_k}{\sigma_k^{2}}\right)^{\delta}\lambda_k \beta_K}{\Gamma(N_{t_k})}\!\int_{0}^{\infty}\!\!\frac{\eta(\xi)}{\xi^{2+\delta}}H_{1,1}^{0,1}\left[\!\frac{\xi}{ \beta_k} \!\left|\!\!\begin{array}{ccc} (2\!-\!N_{t_k},1) \\ (1,1) \end{array}\right. \!\!\!\!\right]  d\xi,\nonumber \\
\label{CMIMO}
\end{eqnarray}
where
\begin{equation}
\eta(\xi)= {\mathcal  H}_{1,1}^{1,1}\!\!\left(\sum_{j=1}^{{\cal T}}\!\!\! \frac{\pi \lambda_j\widetilde{P}^{\delta}_j}{\left(\frac{P_k}{\xi\sigma^{2}_k}\right)^{-\delta}}  \!\left(1\!+\!\delta \xi H_{q+2,p+3}^{v+1,u+2}\!\!\left(\xi, {\cal P}^{{\cal I}_j}_{\cal U}\right)\!\right)\!,\!{\cal P}_{\delta}\!\!\right).
\label{eta}
\end{equation}
 Compared with existing approaches in  \cite{lee}-\!\!\!\cite{vm4}, which requires the calculation of  $N_{t_k}-1$ derivatives of $\eta(\xi)$ when $g_{x_k}$ is gamma distributed as $\texttt{Gamma}(N_{t_k},1)$, the framework in (\ref{CMIMO}) and (\ref{eta}) adds no computational complexity
and thus preserves the tractability of single-antenna settings.
We note that assuming a Gamma distribution for the interferers' power gain, i.e. $g_{x_j}\sim\texttt{Gamma}(\chi_j, \phi_j)$, $j\in \{1,\ldots, {\cal T}\}$, is  commonly encountered in multi-antenna networks \cite{xu}, \cite{xu1}, \cite{deng}.  In this case,  we only need to modify the parameters of  $\eta(\xi)$  by replacing in (\ref{CMIMO}) the following equation
 \begin{eqnarray}
{\cal P}^{{\cal I}, j}_{\cal U}\!\!\!\!\!&=&\!\!\!\!\!\!\bigg(\frac{\phi_j}{\Gamma(\chi_j)}, \phi_j, (-\chi_j,0,\delta),(0, 1, ,\delta\!-\!1),\nonumber \\ && (1,1,1), (1,1,1,1)\bigg).
\end{eqnarray}

\subsubsection{Multi-Antenna Ad Hoc Networks}
The coverage probability of ad hoc networks  for different  multi-antenna  transmission strategies for which   $g_{x_k}\sim\texttt{Gamma}(M_k, \theta_k)$, $k=1,\ldots, {\cal T}$ is directly obtained from  (\ref{outcu}) as
\begin{eqnarray}
\!\!\!{\cal C}&=&\sum_{k=1}^{{\cal T}}\frac{1}{\Gamma(M_k)} \nonumber \\&& \!\!\!\!\!\!\!\!\!\!\!\!\!\!\!\!\!\!\times H_{1,2}^{2,0}\left[\sum_{j=1}^{{\cal T}}\frac{\pi r_k^{2} \lambda_j \Gamma(1-\delta)\Lambda_j  \beta_k^{\delta} }{\widetilde{P}^{-\delta}_j\left(\theta_k\right)^{-\delta}}\left|
\begin{array}{ccc}(1,\delta) \\ (0,1), (M_k, \delta ) \end{array}\right.\right],
\label{outcucsdma}
\end{eqnarray}
where $\Lambda_j$ accounts for Fox's H identically distributed interferences and is given in (\ref{lambdaj}).
 The coverage probability scaling law of multi-antenna ad hoc networks using MRT with $N_{t_k}$ antenna at the $k$-th tier BS is obtained  from  applying (\ref{as1}) to (\ref{outcucsdma}) as
\begin{eqnarray}
{\cal C}&\underset{\lambda\rightarrow\infty}{\approx}& \!\!\sum_{k=1}^{{\cal T}} \frac{\left(\lambda {\cal A} \right)^{N_{t_k}-1}}{\Gamma(N_{t_k})}\exp\left( -\lambda {\cal A} \right)\underset{\lambda\longrightarrow\infty}{\longrightarrow} 0
\label{asmimo1}
\end{eqnarray}
where ${\cal A}=\sum_{j=1}^{{\cal T}}\!\! \frac{\pi r_k^{2}  \Gamma(1\!-\!\delta)\Lambda_j }{\widetilde{P}^{-\delta}_j  \beta_k^{-\delta}}$. This last result reveals that, although the SIR increases in multi-antenna ad hoc networks, it will continue to drop to zero as the transmitter density increases.

\subsection{The Impact of the Antenna Size}
In this subsection,  we consider multiple-input-single-output single-tier networks (i.e., ${\cal T}=1$) in which the BSs are equipped with $N_t$ antennas. Next we  exploit the expressions and tools of the previous sections to derive the scaling laws for
 different multi-antenna networks including ad hoc, cellular, mmWave and  networks with elevated BSs.

\subsubsection{Antenna Scaling in Ad Hoc Networks}
 For the multi-antenna case,  the coverage expression in
 (\ref{outcucsdma}) can be used to find the asymptotic scaling laws summarized as follows.

\textit{Proposition 5:} Consider a  multiple-input-single-output  ad hoc network with $N_t$ transmit antennas such that $\underset{\lambda\rightarrow\infty}{\lim} \frac{N_t}{\lambda^{\frac{1}{\delta}}}= \gamma$, where $\gamma\in [0, \infty]$,  then the asymptotic coverage probability  has the following scaling law
\begin{equation}
\underset{\lambda\rightarrow\infty}{\lim}{\cal C}=\left\{
                                                    \begin{array}{ll}
0, & \hbox{$\gamma=0$;} \\

                                                     H_{1,1}^{1,0}\left[ \frac{T}{\gamma^{\delta}}\left|
\begin{array}{ccc}(1,\delta) \\ (0,1) \end{array}\right.\right], & \hbox{$\gamma\in\mathbb{R}^{*}_{+}$;} \\

                                                      1, & \hbox{$\gamma=\infty$.}
                                                    \end{array}
                                                  \right.
\end{equation}
where $\gamma=0, \in\mathbb{R}^{*}_{+}, \infty$ stands for asymptotically  sublinear, linear and super-linear scaling of $N_t$ and $T= \pi r^{2} \Gamma(1\!-\!\delta) \beta^{\delta} \theta^{\delta}\Lambda$.

 \textit{Proof:}  Resorting to the  Mellin-Barnes representation of the Fox's H-function in (\ref{outcucsdma}), it follows that
\begin{eqnarray}
C&\overset{(a)}{=}&\frac{1}{2\pi j} \int_{\cal C}\frac{ \Gamma(N_t+\delta s)\Gamma(s)}{\Gamma(N_t)\Gamma(1+\delta s)} \left(T\lambda\right)^{-s}ds\nonumber \\
&\underset{\lambda\rightarrow\infty}{\overset{(b)}{\simeq}}&\frac{1}{2\pi j} \int_{\cal C}\frac{\left( T\frac{\lambda}{N_t^{\delta}}\right)^{-s}}{\Gamma(1+\delta s)} ds,
\label{eq1}
\end{eqnarray}
where  $(a)$ follows  from using  \cite[Eq. (2.1)]{mathai} and  $(b)$ follows form applying  $\underset{\lambda\rightarrow\infty}{\lim }\frac{\Gamma(N_t+\delta s)}{\Gamma(N_t)} =(N_t)^{\delta s}$.  The proof follows by recognizing the Fox's H function definition in \cite[Eq. (1.2)]{mathai}, along with its asymptotic expansions near zero \cite[Eq. (1.7.14)]{kilbas} and infinity \cite[Eq. (1.8.7)]{kilbas}.

Hence, based on  Proposition 5, we  evince that scaling the number of antennas linearly with the density does not prevent the SINR from
dropping to zero for high BSs densities (as $\delta^{-1}=\alpha/2$  with $\alpha>2$) thereby hindering the SINR invariance property. Interestingly, when the number of antennas scales  super-linearly
with the BSs density, the coverage approaches a finite
constant which is desirable since it guarantees a certain quality of service (QoS) for the users in the dense regime.
\subsubsection{Antenna Scaling in Cellular Networks}

Before delving into the analysis, it is important to recall that, in the single antenna case, the
coverage probability under a practical bounded path-loss model drops to zero as $\lambda \rightarrow\infty$ (see Section II.C).
In the multi-antenna case,  the asymptotic coverage scaling laws are summarized in the following proposition.

\textit{Proposition 6:} In multiple-input-single-output  cellular networks with $N_t$ transmit antennas such that $\underset{\lambda\rightarrow\infty}{\lim} \frac{N_t}{\lambda}= \zeta$, where $\zeta\in [0, \infty]$, the asymptotic coverage probability under a  bounded path-loss model has the following scaling law
\begin{equation}
\underset{\lambda\rightarrow\infty}{\lim}{\cal C}=\left\{
                                                    \begin{array}{ll}
0, & \hbox{$\zeta=0$;} \\
                                                   \!\!  H_{1,1}^{1,0}\!\left[\!2\pi  \frac{  \beta}{\eta \zeta}\!\left|\!
\begin{array}{ccc} (1,1) \\ (0,1) \end{array}\right.\right], & \hbox{$\zeta\in\mathbb{R}^{*}_{+}$;} \\

                                                      1, & \hbox{$\zeta=\infty$,}
                                                    \end{array}
                                                  \right.
\end{equation}
where $\eta= 2-3 \alpha+\alpha^{2}$ and $\zeta=0, \in\mathbb{R}^{*}_{+}, \infty$ stands for asymptotically sub-linear, linear and super-linear scaling of $N_t$.

\textit{Proof:}
 Due to the intricacy of ${\cal L}_{I}$ in (\ref{IL2}), we resort to an analytically tractable tight lower bound. Under the bounded path-loss model, the coverage probability in (\ref{p1}) involves the interference Laplace transform  ${\cal L}_{{\cal I}}(s(1+r)^{\alpha}) = \exp(2 \pi \lambda \Theta(s(1+r)^{\alpha})$,
 where  we have \begin{eqnarray}
\Theta(s)&=&E_{g}\left[\int_{r}^{\infty}\left(1-\exp\left(-s g (1+t)^{-\alpha})\right)\right)t dt\right]\nonumber \\
&\overset{(a)}{\leq} & s E[g] \int_{r}^{\infty}  t(1+t)^{-\alpha} dt\nonumber \\
&\overset{(b)}{\approx}& \frac{s}{\eta}\left((1+ r)^{1-\alpha} (1-r+\alpha r)\right),
\label{eqa1}
\end{eqnarray}
where the inequality in $(a)$ follows from the fact that $1-e^{-x}\leq x$, $\forall x\geq0$ and  $(b)$ holds since $g$  has a unit mean. Note that when $r$ becomes smaller, the inequality
in (\ref{inq}) becomes tighter. This is typically the case in  ultra-dense networks, where the closest distance to the
origin tends to be infinitesimally small. Accordingly, by relabeling $r\leftarrow\frac{r}{\lambda}$, we obtain
\begin{eqnarray}
\Theta\left(s\left(1+\frac{r}{\lambda}\right)^{\alpha}\right)  &\underset{\lambda\rightarrow\infty}{\approx}&\frac{s}{\eta}.
\label{inq}
\end{eqnarray}
 Hence, the coverage probability can be obtained by merging (\ref{inq}) and (\ref{p1})  as
\begin{eqnarray}
{\cal C}&\approx& \frac{ \beta}{\Gamma(N_t)}\int_{0}^{\infty}\!\!\frac{e^{- 2\pi \frac{\lambda }{\eta }\xi}}{\xi^{2}}H_{1,1}^{0,1}\!\!\left[ \!\frac{ \xi}{ \beta} \left|\!\!\!\begin{array}{ccc} (2-N_t,1) \\ (1,1) \end{array}\right. \!\!\!\right]  d\xi,\nonumber \\
&\overset{(a)}{\approx}&  \frac{ 1}{\Gamma(N_t)}H_{1,2}^{2,0}\!\!\left[ 2\pi \frac{\lambda \beta}{\eta } \left|\!\!\!\begin{array}{ccc}(1,1) \\ (0,1), (N_t,1) \end{array}\right. \!\!\!\right],
\label{Csb}
\end{eqnarray}
where $(a)$ follows from applying $\int_{0}^{\infty}f_r(r) dr=1$ and \cite[Eq. (2.3)]{mathai}.  As $N_t(\lambda)\underset{\lambda\rightarrow\infty}{\rightarrow} \infty$, we obtain
\begin{equation}
C \underset{\lambda\rightarrow\infty}{\overset{(b)}{\simeq}}H_{1,1}^{1,0}\left[ 2\pi \frac{\lambda \beta}{ N_t \eta } \left|\begin{array}{ccc}(1,1) \\ (0,1) \end{array}\right. \right],
\label{cc1}
\end{equation}
where $(b)$ follows along the same lines of (\ref{eq1}). The proof follows by resorting to the asymptotic expansions of the Fox' H function  in (\ref{Csb}) when $\zeta=\frac{N_t}{\lambda}$ is near zero \cite[Eq. (1.7.14)]{kilbas} and infinity \cite[Eq. (1.8.7)]{kilbas}.

The obtained result in (\ref{cc1}) allows us to conclude that monotonically increasing the per-user coverage performance by means of ultra-densification is theoretically possible through the deployment of multi-antenna BSs. Specifically, (\ref{cc1}) unveils  that scaling linearly the number of antennas with the BS density constitutes a solution for the coverage drop  in traditional dense networks.

\subsection{ The Impact of Antenna Gain in mmWave Networks}
In multiple-input-single-output mmWave networks, the channel gain for the signal $g_{x}$ follows a
gamma distribution $g_{x}\sim\texttt{Gamma}(N_t, \frac{1}{N_t})$, where $N_t$ is the number of antennas at the BS \cite{xu}-\!\!\cite{xu1}. As for the interference received at the typical user,  the total channel gain  is the product of an arbitrary unit mean small-scale fading gain $g$ \cite{renzo2},  \cite{xu1} and the directional antenna array gain $G(\frac{d}{\lambda_t} \theta_x)$,  where $d$ and $\lambda_t$  are the antenna spacing and  wavelength, respectively,  and $\theta_x$ is a
uniformly distributed random variable over $[-1,1]$. An example of  antenna pattern based on the cosine
function is given by \cite{xu1}, \cite{deng}
\begin{equation}
G(x)=\left\{
        \begin{array}{ll}
          \cos^{2}\left(\frac{\pi N_t}{2} x\right), & \hbox{ $|x|\leq \frac{1}{N_t}$;} \\
          0, & \hbox{otherwise.}
        \end{array}
      \right.
\end{equation}

In dense mmWave network deployments, it is reasonable to assume that the
link between any serving BS and the user is  in line-of-sight (LOS). Mathematically, the probability of being in a LOS propagation
can be formulated as $p(r)=e^{-\tau r}$, where $\tau$ is the blockage parameter determined by the density and
average size of the spatial blockage \cite{atzeni}, \cite{xu}.  Accordingly, (\ref{eqa1}) can be derived, based on the cosine antenna pattern and the blockage model, as
\begin{eqnarray}
\!\!\Theta(s) \!\!\!\!\!&\leq& \!\!\!\!\!\!\!\!\frac{\lambda_t s}{\pi d N_t}\left( \int_{r}^{\infty}\!\!\!\! t \frac{e^{-\tau t}}{(1+t)^{\alpha_L}} dt \!+\! \int_{(1\!+\!r)^{\frac{\alpha_L}{\alpha_N}}\!-\!1}^{\infty}\!\!\frac{t(1\!-\!e^{-\tau t})}{(1+t)^{\alpha_N} }dt\right)\nonumber \\ && \times \int_{0}^{\pi} \cos^{2}\left(\frac{x}{2}\right)dx
\nonumber \\ &\underset{\lambda\rightarrow\infty}{\overset{(a)}{=}}&\frac{\lambda_t e^{\tau} s}{2 d N_t}\left({\cal P}+{\cal J}(\tau)\right),
\label{inq1}
\end{eqnarray}
where $(a)$ follows along the same lines of (\ref{inq}) with $\alpha_L(\alpha_N)$ is the path-loss exponent of the LOS (NLOS) link, ${\cal P}=\frac{\alpha_N-\alpha_L-1}{(1-\alpha_L)(\alpha_N-2)}$, ${\cal J}(\tau)=\frac{(\alpha_L-1+\tau)E_{\alpha_L-1}(\tau)}{\alpha_L-1}-\frac{(\alpha_N-1+\tau)E_{\alpha_N-1}(\tau)}{\alpha_N-1}$, with $E_{\nu}(\cdot)$ denoting the Exponential Integral function \cite{grad}. Then, using (\ref{inq1}) and  following  the same steps as in (\ref{Csb}), we obtain
\begin{eqnarray}
\!\!\!\!\!\!\!{\cal C}\!\!\!
&\approx&\!\!\!\!\!\frac{ 1}{ \Gamma(N_t)}H_{1,2}^{2,0}\!\!\left[ \! \frac{\pi \lambda\lambda_t e^{\tau} \beta}{ d}\left({\cal P}\!+\!{\cal J}(\tau)\right)\! \left|\!\!\!\begin{array}{ccc}(1,1) \\ (0,1), (N_t,1) \end{array}\right. \!\!\!\!\right].
\label{Csb1}
\end{eqnarray}
Hence, the  coverage scaling laws in  mmWave networks  are given in the following  proposition.

\textit{Proposition 7:} In mmWave networks in which  $\lim_{\lambda\rightarrow\infty}\lambda_t=0$, and  $\underset{\lambda\rightarrow\infty}{\lim} \frac{Nt \lambda_t^{-1} }{ \lambda}= \rho$, where $\zeta\in [0, \infty]$,  the asymptotic coverage probability has the following scaling law
\begin{equation}
\underset{\lambda\rightarrow\infty}{\lim}{\cal C}=\left\{
                                                    \begin{array}{ll}
0, & \hbox{$\rho=0$;} \\
                                                   \!\!  H_{1,1}^{1,0}\!\!\left[ \pi \frac{ \beta e^{\tau}\left({\cal P}+{\cal J}(\tau)\right) }{ d \rho} \left|\!\!\!\begin{array}{ccc}(1,1) \\ (0,1) \end{array}\right. \!\!\!\right], & \hbox{$\rho\in\mathbb{R}^{*}_{+}$;} \\

                                                      1, & \hbox{$\rho=\infty$.}
                                                    \end{array}
                                                  \right.
\end{equation}

\textit{Proof:} The proof is similar to those of  Propositions 5 and 6.\\
The obtained result unveils that the scaling laws derived for mmWave cellular networks are similar to those obtained for  legacy frequency bands (see Proposition 6). Specifically, maintaining a  linear scaling between the density of BSs and the number of antennas is sufficient to prevent the SINR from dropping to zero and to guarantee a certain QOS to the UE. In addition, the optimal coverage can be achieved by linearly scaling the number of antennas and the  mmWave carrier frequency, which  reduces both cost and power consumption. This result provides evidence that moving toward higher frequency bands may be an attractive solution for high capacity ultra-dense networks.

Achieving optimal coverage rely on  determining the optimal scaling factor below which further densification becomes destructive
or cost-ineffective. This operating point will depend on
properties of the channel power distribution and pathloss and is of cardinal
importance for the successful deployment of  ultra-dense networks.

\textit{Corollary 1 (Optimal Scaling Factor in Dense mmWave Networks):} Capitalizing on Proposition 7, the optimal scaling factor $\rho$  that prevents the outage drop in dense mmWave networks  is given by
\begin{eqnarray}
\rho&=&\frac{N_t f_c}{ \lambda}\nonumber \\
&\overset{(a)}{=}& \frac{\pi \beta e^{\tau}\left({\cal P}+{\cal J}(\tau)\right) }{ d },
\label{op1}
\end{eqnarray}
where $f_c$ is the mmWave carrier frequency and  $(a)$ follows from recognizing that $H_{1,1}^{1,0}\!\!\left[ x\left|\!\!\!\begin{array}{ccc}(1,1) \\ (0,1) \end{array}\right. \!\!\!\right]=U(1-x)$, where $U(x)=\left\{
                                                              \begin{array}{ll}
                                                                1, & \hbox{ $ x\geq0$;} \\
                                                                0, & \hbox{otherwise.}
                                                              \end{array}
                                                            \right.$ stands for the Heaviside function \cite{grad}. In particular, (\ref{op1}) unveils that under  a full-blockage scenario (i.e., $\tau\rightarrow\infty$), a super-linear scaling of $N_t f_c$  is required to offset the coverage drop.  However, in the no-blockage regime (i.e., $\tau\rightarrow 0$), only a linear  scaling is needed. Using this framework,  enhanced antenna
models can be considered  to investigate the impact of beam alignment errors on the coverage probability of mmWave dense networks \cite{cheng}.

\subsection{The  Impact of Antenna Height in 3D Networks}
The vast majority of spatial models for cellular networks are usually 2D and ignore the impact of the BS height. Recent papers have, however, tackled this issue and have highlighted the importance of taking this parameter into account to appropriately estimate the network performance \cite{ding}, \cite{atzeni},\cite{filo}. In 3D cellular networks, the distance
between a BS and the typical UE can be expressed as $\sqrt{h^{2}+r^{2}}$, where $h$ is the absolute antenna height difference between the serving
BS and the typical UE. Adapting the coverage probability expression in (\ref{p1})  to the 3D context results in an interference distribution whose Laplace transform is of the form  ${\cal L}_{{\cal I}}(s(r^{2}+h^{2})^{\frac{\alpha}{2}}) = \exp(2 \pi \lambda \Theta(s(r^{2}+h^{2})^{\frac{\alpha}{2}}))$  where
\begin{eqnarray}
\Theta(s) &\leq& s E[g] \int_{r}^{\infty}  t(h^{2}+t^{2})^{-\frac{\alpha}{2}} dt \\ \nonumber  &\overset{(a)}{=}& \frac{s}{\alpha-2}\left(h^{2}+ r^{2}\right)^{1-\alpha/2}.
\label{inq22}
\end{eqnarray}
By employing the change of variable $x\leftarrow \lambda r^{2}$, we obtain ${\cal L}_{{\cal I}}\left(s\left(\frac{r}{\lambda}+h^{2}\right)^{\frac{\alpha}{2}}\right)\underset{\lambda\rightarrow\infty}{\approx} e^{ \frac{ 2\pi \lambda s h^{2}}{\alpha-2}}$. Hence, the coverage probability in 3D multi-antenna cellular networks  can be formulated similar to (\ref{Csb}) and (\ref{cc1}) as
\begin{equation}
C \underset{\lambda\rightarrow\infty}{{\simeq}} H_{1,1}^{1,0}\left[ 2\pi \frac{\lambda h^{2} \beta}{ N_t (\alpha-2) } \left|\begin{array}{ccc}(1,1) \\ (0,1) \end{array}\right. \right].
\label{chei}
\end{equation}
The obtained analytical expression for the coverage probability unveils the impact of the antenna height coupled with other design parameters.
Recent works  \cite{ding}-\!\!\cite{filo} proposed to maintain the SINR invariance of the coverage probability by lowering  the height of the BSs. Based on (\ref{inq22}), we evince that the SINR invariance of the coverage probability  in 3D networks can be maintained  by enforcing a super-linear scaling with the number of antennas.

\textit{Corollary 2 (Optimal Scaling Factor in Dense 3D Networks):} The optimal scaling factor for the successful deployment of dense 3D networks is
\begin{equation}
\frac{N_t}{ \lambda}= \frac{ 2\pi \beta h^{2} }{ \alpha-2 },
\label{s1}
\end{equation}
which exploits (\ref{chei}) and  follows along the same lines of Corollary 1.  In particular,  the last result shows that the coverage  probability monotonically decreases as the BS density increases, if $\underset{\lambda\rightarrow\infty}{\lim} \frac{Nt}{\lambda}= \zeta \in \mathbb{R}^{*}_{+}$, and if $h>\sqrt{\frac{(\alpha-2 )\zeta}{2\pi }}$.  Interestingly, it is possible to counteract this decay by tuning the antenna number
 according to BS density in order to maintain the per-user coverage performance.

\section{Numerical Results}
In this section, we substantiate our theoretical coverage expressions and scaling laws using system level simulations.  Unless otherwise stated, the noise power is set to $\sigma^{2}=-70$ dBm and the path loss is given by $L(r)=r^{-\alpha}$ for power-law  unbounded model and $L(r)=(1+r)^{-\alpha}$ for physically feasible bounded model, with $\alpha=3$.

The performance comparisons between strongest-BS- and closest-BS-association-based two-tier (i.e., ${\cal T}=2$) cellular networks with unbounded power-law path-loss model are illustrated in Fig.~1.  Overall, the strongest-BS strategy provides significant
performance gain over the closest-BS strategy  especially in low density range. Furthermore, depending
on the target SINR thresholds, the effect of increasing densification is beneficial while, in some cases,  tends to be negligible. Indeed, since using the power law model,  the coverage
saturates to a non-zero finite constant in the limit of $\lambda_1, \lambda_2 \rightarrow\infty$.

Fig.~2 plots the scaling of the coverage probability with  BS
densities for both bounded and  unbounded path-loss models. Analytical and experimental curves are in full agreement.
It shows that the unbounded model (i.e., $r^{-\alpha}$) guarantees a certain QoS
or coverage for the users in the dense regime by preventing the SINR form dropping to zero. However,  this SINR-invariance property is unattainable because the unbounded model  is physically impracticable and unrealistic.  The figure also highlights the diminishing gains achieved with the more realistic bounded, as anticipated by Eq. (16). In this case, new densification strategies are  required to prevent the SINR from dropping to zero and avoid the densification plateau.  This will be discussed later in Fig. 6.

\begin{figure}[h]
\centering
\includegraphics[scale=0.35]{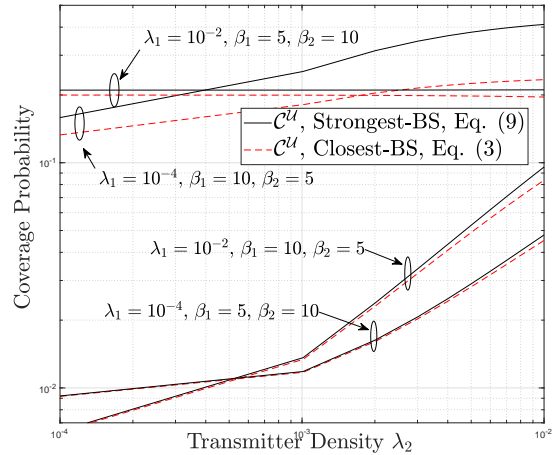}
\caption{Coverage probability   vs. the BS density $\lambda$ for  multi-tier cellular networks with ${\cal T}=2$ over arbitrary Nakagami-$m$ fading with $m_1=1.5$, $m_2=2.5$, $P_1 = 50$ W, and $P_2 = 1$ W.}
\end{figure}
\begin{figure}[h]
\centering
\includegraphics[scale=0.35]{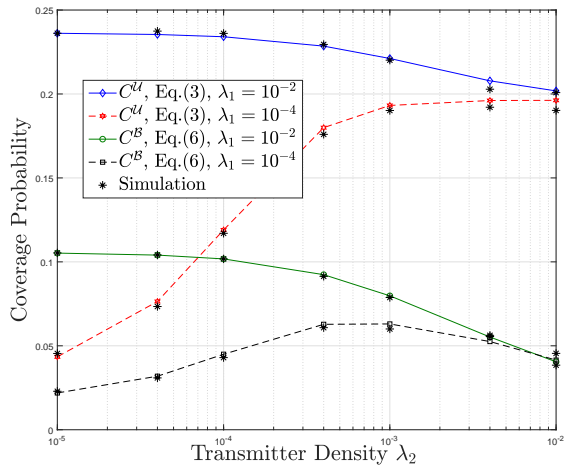}
\caption{Coverage probability and scaling laws vs. the BS density $\lambda$ for  multi-tier cellular networks  over Nakagami-$m$ fading with ${\cal T}=2$, $P_1 = 50$ W, and $P_2 = 1$ W.}
\end{figure}

Fig.~3  shows the scaling of the SIR coverage probability of ad hoc networks against the transmitter
density for various common fading distributions stemming from the general Fox's H fading model. In particular,
we corroborate the result of Eq. (17) stipulating  that increasing the transmitter density degrades the coverage probability in ad hoc
networks, and that the coverage probability is a product of an exponential function and a polynomial
function of order ${\cal T}(m-1)$ of the transmitter density. Moreover, the multi-path fading model has a less noticeable impact on the coverage
performance than the path-loss model (cf. Fig.~2) and the number of tiers.

Fig. 4 shows the SIR outage probability of cellular  networks for an unbounded path-loss model versus the antenna size when assuming that the  interferers' power gain follows a Gamma distribution, i.e., $g\sim\texttt{Gamma}(\chi, \phi)$.  Fig. 4 demonstrates that increasing the antenna size keeps improving the coverage probability, less significantly, however,  as the number of antennas grows large.

\begin{figure}[h]
\centering
\includegraphics[scale=0.35]{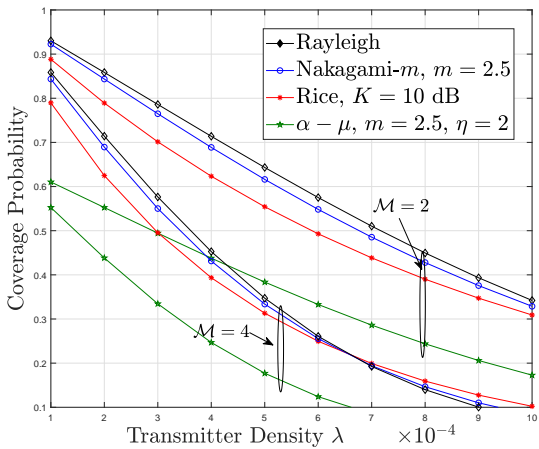}
\caption{Coverage probability in ad hoc networks vs.  the transmitter density $\lambda$  when $\beta=0$ dB.}
\end{figure}
\begin{figure}[h]
\centering
\includegraphics[scale=0.35]{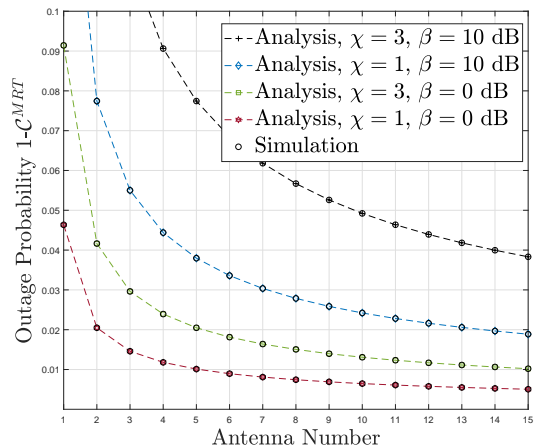}
\caption{Outage probability in MISO cellular networks assuming MRT vs. the number of antennas at the transmitter with $\lambda=10^{-3}$ and $\phi=1$.}
\end{figure}
\begin{figure}[h]
\centering
\includegraphics[scale=0.35]{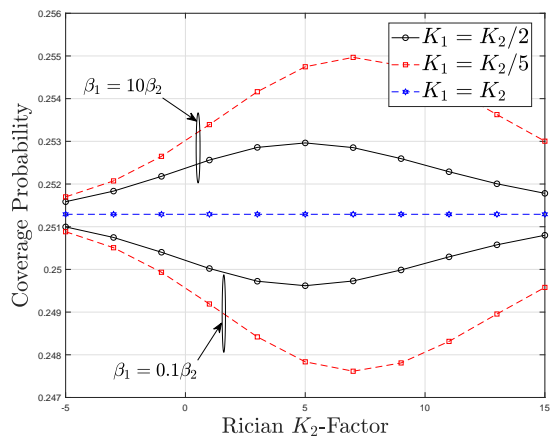}
\caption{Coverage probability in a two-tier cellular network under strongest-BS association vs. the Rician K-factor for  $\lambda=10^{-3}$.}
\end{figure}

 Fig.~5  illustrates the SIR coverage probability of a two-tiers cellular network over Rician fading with closed-BS association obtained from Proposition 3 for different Rician power factors. We observe a substantial increase of the coverage probability only in the non-asymptotic regime, i.e., $K_1\neq K_2$.   Moreover, we observe that the two extreme regimes of
pure fading channel with ($K\rightarrow0$) and pure LOS propagation ($K\rightarrow\infty$)  achieve worse coverage performance.

\begin{figure}[h]
\centering
\includegraphics[scale=0.35]{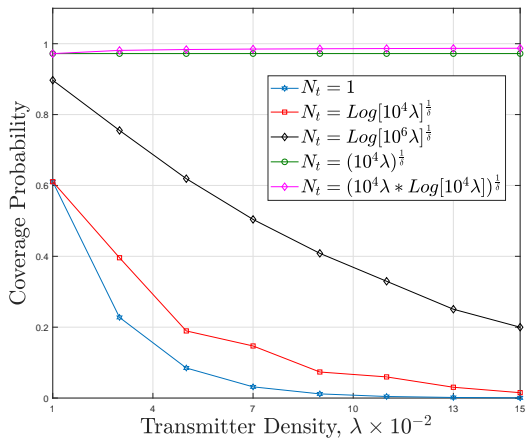}
\caption{Coverage probability  of MISO ad hoc networks vs the BS density $\lambda$ for different scaling of the number of antennas. }
\end{figure}

Fig. 6 shows the scaling of the coverage probability in ad hoc networks against the transmitter
density for different scaling rates of the number of antennas;
super-linear, linear, sub-linear, and constant (i.e, single antenna).
We notice  that the coverage decreases with
the density for the single antenna case, as anticipated in (\ref{asmimo1}), and also when the number of antennas is scaled sublinearly or linearly with the density, as predicted by Proposition 5.
We observe that  a super-linear scaling of the number of
antennas with the BS density is required to prevent the SIR
from dropping to zero, and thereby restore the SIR invariance property.
\begin{figure}[h]
\centering
\includegraphics[scale=0.35]{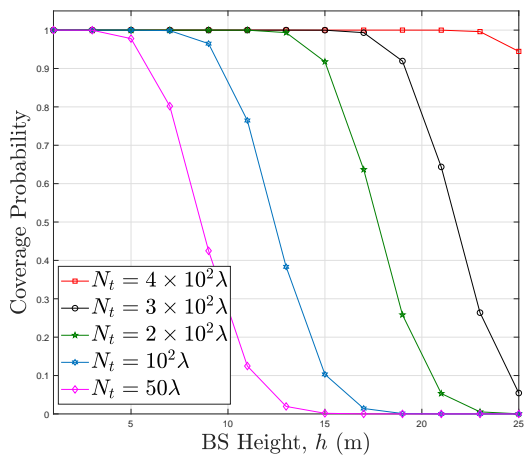}
\caption{Coverage probability  of MISO cellular networks vs the BS height $h$ for different scaling of the antenna number $N_t$. }
\end{figure}

The impact of the BS height on the coverage probability is illustrated in Fig.~7. As predicted in Section IV.D, we note that a linear scaling of the number of antennas is required to maintain a non-zero SINR for low value of $h$. When the BS height increases, the coverage probability decreases due to the increase of the path-loss and the linear scaling becomes insufficient.

%

\section{Conclusion}

By leveraging the properties of Fox's H random variables, we developed
a unifying  framework to characterize the coverage probability of heterogeneous and muti-antenna networks
 under both the closest-BS and the strongest-BS cell association strategies. We studied the impact of BS densification on the coverage performance  both  under bounded and unbounded path loss models. By direct inspection of the obtained analytical framework, we have been able to derive  exact closed-form formulations and scaling laws of the coverage probability  for two typical network models, i.e., heterogeneous and multi-antenna cellular and ad hoc networks, while  incorporating generalized fading distributions. The obtained results encompass insightful relationships between the BS density and the relative antenna array size, gain  and  height,  showing how the coverage can be
maintained whilst increasing the network density.  The insights provided in this work  are of cardinal importance for optimally deploying  general ultra-dense networks.

\section{Appendix A: Proof of Proposition 1}
With the closest-BS association strategy,  the coverage probability is given by
\begin{equation}
 {\cal C}\triangleq\sum^{{\cal T}}_{k=1}{\cal \theta}_k   {\cal P}\left(\text{SINR}_{k}>\beta_k\right),
  \end{equation}
  where  ${\cal \theta}_k$ denotes the  association probability and  is expressed as ${\cal \theta}_k= \frac{\lambda_k}{\sum_{j\in {\cal T}}\lambda_j \widetilde{P}_j^{\delta}}$, and  $\widetilde{P}_j=\frac{P_j}{P_k}$.
Using  \cite[Theorem 1]{trigui} and \cite[Eq. (39)]{trigui1} and  applying the Fox's $H$-transform in \cite[Eq. (2.3)]{mathai}, the coverage probability under unbounded path-loss model, denoted as ${\cal C}^{\cal U}$, is given by 
\begin{eqnarray}
{\cal C}^{\cal U}(r_k) &=& \int_{0}^{\infty}\frac{1}{\sqrt{\xi}}~ {\cal L}^{-1}\left\{\frac{1}{\sqrt{s}}H_{p,q}^{u,v}\left\{f(t); {\mathcal P}\right\}(s \xi); s; {\cal \beta}_k\right\} \nonumber\\ &&e^{-\sigma^{2}_k\xi \frac{r_k^{\alpha}}{P_k}}
\prod_{j\in {\cal T}}{\cal L}_{{\cal I}_j}\left(\xi\frac{r_k^{\alpha}}{P_k}\right) d\xi,
\label{p1}
\end{eqnarray}
where $f(t)=\sqrt{t}{\mathcal J}_1\left(2\sqrt{s t \xi }\right)$, $H_{p,q}^{u,v}\left\{f(t); {\mathcal P}\right\}(s)$ is the Mellin transform \cite[Eq. (2.3)]{mathai},  ${\mathcal J}_{1}(x)=H_{0,2}^{1,0}\left(\frac{x^{2}}{4};(1,1, \frac{1}{2},-\frac{1}{2},1,1 )\right)$ is the Bessel function of the first kind \cite[Eq. (8.402)]{grad}, and ${\cal L}^{-1}$ is the inverse Laplace transform.
Moreover ${\cal L}_{{\cal I}_j}$, in (\ref{p1}), is the Laplace transform of the aggregate interference from the $j$-th tier  evaluated as
\begin{equation}
{\cal L}_{{\cal I}_j}(s) = \exp(2 \pi \lambda_j \Theta(s)),
\label{IL}
\end{equation}
where
\begin{eqnarray}
\!\!\Theta\mid_{H_j=y} (s) \!\!\!&\overset{(a)}{=}&\!\!\!\int_{\left(\frac{P_j}{P_k}\right)^{\frac{2}{\alpha}}r_k}^{\infty}\!\!\!\!\left(1-\exp\left(-s  y P_j r^{-\alpha}\right)\right)r dr\nonumber\\&&
\!\!\!\!\!\!\!\!\!\!\!\!\!\!\!\!\!\!\!\!\!\!\!\!\!\!\!\!\!\!\!\!\!\!\!\!\overset{(b)}{=} \frac{s y P_j}{\alpha}  \int_{\frac{P_k}{P_j}r_k^{-\alpha}}^{\infty} x^{-\frac{2}{\alpha}} e^{-s y x}{\rm {}_{\!1}F_{\!1}}\left(1,2,s y x\right )dx \nonumber\\&&
\!\!\!\!\!\!\!\!\!\!\!\!\!\!\!\!\!\!\!\!\!\!\!\!\!\!\!\!\!\!\!\!\!\!\!\!\overset{(c)}{=} \frac{ s  y P_j ~{\rm {}_{2}F_{2}}\left(1, -\frac{2}{\alpha}+1; 2;-\frac{2}{\alpha}+2; - s y P_k r_k^{-\alpha}\right) }{\left(\frac{P_k}{P_j}\right)^{\frac{2}{\alpha}-1}r^{(\alpha-2)}\alpha\left(\frac{2}{\alpha}-1\right)} ,
\label{agg}
\end{eqnarray}
where $(a)$ follows  from the probability generating functional \cite{dilhon}, \cite{han}, while  relabeling $x$ as $r_k^{-\alpha}$ and $(1-e^{-x})/x=  e^{-x}{\rm {}_{1}F_{1}}\left(1, 2; x\right)$ is applied in $(b)$,
and $(c)$ follows from  applying $\int x^{\beta-1} e^{- c x}{\rm {}_{1}F_{1}}(a,b, c x)=\frac{x^{\beta}}{\beta}{\rm {}_{2}F_{2}}\left(b-a, \beta, b, \beta+1, -c x \right)$.
 Hence,  we obtain
\begin{eqnarray}
{\cal L}_{{\cal I}_j}(\xi) &=& \exp(-2\pi \lambda_j \mathcal{E}_{H_j}[\Theta(s)]) \nonumber \\ &&
\!\!\!\!\!\!\!\!\!\!\!\!\!\!\!\!\!\!\!\!\!\!\!\!\!\!\!\!\!\!\!\!\!\!\!\!= \exp \left(\!-\pi \delta \lambda_j \frac{\xi r_k^{2-\alpha}}{\left(1-\delta\right)}H_{p,q}^{u,v}\left\{g(t); {\mathcal P}\right\}(\xi)\!\right),
\label{IL1}
\end{eqnarray}
where $g(t)=t ~{\rm {}_{2}F_{2}}\left(1, 1-\delta; 2;2-\delta; - \xi t r_k^{-\alpha}\right)=tH_{2,3}^{1,2}\left(t;{\cal P}_1\right)$, ${\cal P}_1=(1-\delta, \xi (r_k^{2})^{-\frac{\alpha}{2}}, (0,\delta), (0,-1,\delta-1), {\bf 1}_2, {\bf 1}_3 )$, and ${\rm {}_{p}F_{q}}(\cdot)$ is the generalized  hypergeometric function of \cite[Eq. (9.14.1)]{grad}.
In (\ref{IL1}), in particular,  we first take the expectation over the interferers' locations and then average over the Fox's H distributed channel gains, which is in the reverse order compared to the conventional derivations in
\cite{and}-\!\!\cite{dilhon}. The reason behind this order swapping is that  the Fox's H fading model is  more complicated than the conventional exponential model, and therefore  averaging over it in a latter step preserves the analytical tractability.

Finally, applying \cite[Eq. (1.58)]{mathai}, the  Mellin transform \cite[Eq. (2.3)]{mathai},  and the inverse Laplace transform of the Fox's $H$-function  \cite[Eq. (2.21)]{mathai} given by
\begin{equation}
{\cal L}^{-1}\{x^{-\rho}H_{p,q}^{u,v}(x;{\cal P}); x;t\}=t^{-\rho-1}H_{p+1,q}^{u,v}\left(\frac{1}{t};{\cal P}_l\right),
\label{laplace}
\end{equation}
where ${\cal P}_l=(\kappa, c, (a,\rho), b, (A,1), B)$,  the desired result is obtained  after applying the Fox's H reduction formulae in \cite[Eq. (1.57)]{mathai}.
The  coverage probability over Fox's $H$-fading\!\footnote{We dropped the index $i$ from Fox's $H$-distribution $\{{\cal O}_i, {\mathcal P}_i\}$ for notation simplicity.} for a receiver connecting to a $k$-th tier BS located at $x_k$ is given by
\begin{eqnarray}
{\cal C}^{\cal U}(r_k)\!\!&=&\!\!\int_{0}^{\infty}\frac{1}{\xi^{2}}H_{q,p+1}^{v,u}\left(\xi; {\cal P}^{k}_{\cal U}\right)e^{-\sigma^{2}_k\xi \frac{r_k^{\alpha}}{P_k}}\nonumber\\ &&\!\!\!\!\! \!\!\!\!\!\!\! \!\!\!\!\! \!\!\!\!\!\!\!\!\!\!\!\!\!\!\exp\Bigg(\!\!\!- \pi \delta \sum_{j\in {\cal T}}r_k^{2} \lambda_j \widetilde{P}_j^{\delta} \xi H_{q+2,p+3}^{v+1,u+2}\left(\xi; {\cal P}^{{\cal I}}_{\cal U}\right)\Bigg)d\xi,
\label{cux}
\end{eqnarray}
where $\widetilde{P}_j=\frac{P_j}{P_k}$, $\delta=\frac{2}{\alpha}$,  and the parameter sequences  ${\cal P}^{k}_{\cal U}=\!\!\left(\kappa {\cal \beta}_k, \frac{1}{c {\cal \beta}_k}, 1\!-\!b, (1\!-\!a, 1), \mathcal{B}, (A,1)\right),
$ and ${\cal P}^{{\cal I}}_{\cal U}=\bigg(\frac{\kappa}{c^{2}}, \frac{1}{c}, (1\!-\!b\!-\!2B,0,\delta),(0, 1\!-\!a\!-\!2A, -1,\delta\!-\!1),$ $ (\mathcal{B},1,1), (1,A,1,1)\bigg)$.
 Recall   that the pdf  of the link's distance $r_k$  is given by $f_{r_k}(x)=\frac{2\pi  \lambda_k}{\theta_k}  x \exp\left(-\sum_{j\in {\cal T}}\pi x^{2} \lambda_j \widetilde{P}_j^{\delta}\right)$ \cite{and3}, \cite{han}. Then recognizing that $\exp(-x)=H_{0,1}^{1,0}(x;{1,1, 0, 1, 1,1})$ \cite[Eq. (1.125)]{mathai} in (\ref{cux}), we apply \cite[Eq. (2.3)]{mathai} to obtain the average coverage probability  in (\ref{cb}) after some manipulations.

\section{Appendix B: Proof of Proposition 2}
The proof of Proposition 2 relies on the  same approach adopted in Appendix A, yielding
\begin{eqnarray}
{\cal C}^{\cal B}(r_k)\!\!\!\!&=&\!\!\!\! \int_{0}^{\infty}\frac{1}{\sqrt{\xi}} ~ {\cal L}^{-1}\left\{\frac{1}{\sqrt{s}}H_{p,q}^{u,v}\left\{f(t); {\mathcal P}\right\}(s \xi); s; {\cal \beta}_k\right\}\nonumber\\ && e^{-\sigma^{2}_k\xi \frac{(1+r_k)^{\alpha}}{P_k}}
\prod_{j\in {\cal T}}{\cal L}_{{\cal I}_j}\left(\xi\frac{(1+r_k)^{\alpha}}{P_k}\right) d\xi,
\label{p1}
\end{eqnarray}
where rearranging \cite[Eq. (39)]{trigui1} after carrying out the  change of variable relabeling $(1+x)^{-\alpha}$ as $x$, we have
\begin{eqnarray}
\!{\cal L}_{{\cal I}_j}(\xi)\!\!\!\! &=& \!\!\!\!\exp \!\!\Bigg(\!\!\!-\pi \delta \lambda_j \xi \Bigg(\!\!\frac{(1\!+\!r_k)^{2-\alpha}}{\left(1-\delta\right)}H_{p,q}^{u,v}\!\!\left\{g_1(t); {\mathcal P_1}\right\}(\xi)-\nonumber\\ &&\frac{(1+r_k)^{1-\alpha}}{\left(1-\frac{\delta}{2}\right)}H_{p,q}^{u,v}\left\{g_2(t); {\mathcal P}\right\}(\xi)\Bigg)\Bigg),
\label{IL2}
\end{eqnarray}
where $g_1(t)=t ~{\rm {}_{2}F_{2}}\left(1, 1-\delta; 2; 2-\delta; - \xi t (1+r_k)^{-\alpha}\right)$ and $g_{2}(t)=t ~{\rm {}_{2}F_{2}}\left(1, 1-\frac{\delta}{2}; 2;2-\frac{\delta}{2}; - \xi t (1+r_k)^{-\alpha}\right)$.
Finally  applying the Mellin transform in  \cite[Eq. (2.3)]{mathai} and  plugging the obtained result
into (\ref{p1}), Proposition 2 follows after some manipulations.

\section{Appendix C: Proof of Proposition 3}
Based on \cite{trigui1},  the Laplace transform of the aggregate interference from tier
$j$ under the max-SINR association strategy is evaluated as ${\cal L}_{{\cal I}_j}(\xi)= \exp\left(- \pi  \lambda_j \xi^{\delta}\Gamma\left(1-\delta\right)  {\cal E}[H^{\delta}]\right)$, where
${\cal E}[H^{\delta}]$ is the Mellin transform of the Fox's-$H$ function obtained as  ${\cal E}[H^{\delta}]=\Lambda$ \cite[Eq. (2.8)]{mathai}. Then following the analytical steps as in  Appendix A, we obtain
\begin{eqnarray}
{\cal C}&=&  \sum_{k\in {\cal T}}\lambda_k{\cal E}_{r_k} \Bigg\{\int_{0}^{\infty}\frac{1 }{\xi^{2}}H_{q,p+1}^{v,u}\left(\xi; {\cal P}^{k}_{\cal U}\right)\nonumber\\ &&\exp\Bigg(-\sum_{j\in {\cal T}}r_k^{2} \pi \lambda_j \widetilde{P}^{\delta}_j \xi^{\delta}\Gamma(1-\delta)\Lambda_j\Bigg)d\xi\Bigg\}\nonumber \\
&\overset{(a)}{=}&\sum_{k\in {\cal T}}\frac{\lambda_k}{\delta}{\cal E}_{r_k} \Bigg\{\int_{0}^{\infty}\frac{1 }{\xi^{2}}H_{q,p+1}^{v,u}\left(\xi; {\cal P}^{k}_{\cal U}\right)\nonumber\\ &&\!\!\!\!\! \!\!\!\!\!\!\!\!\!\!\!\!\!\!H_{0,1}^{1,0}\!\!\left(\!\!\left(\!\sum_{j\in {\cal T}}r_k^{2} \pi \lambda_j \widetilde{P}^{\delta}_j \Gamma(1\!-\!\delta)\Lambda_j\!\!\right)^{\!\frac{1}{\delta}}\!\!\!\xi;\left(1,1, 1, 0, 1,\frac{1}{\delta}\right)\!\!\right)\!\!d\xi\!\Bigg\},\nonumber\\
\label{cm}
\end{eqnarray}
where $(a)$ follows from substituting $\exp(-x)=H_{0,1}^{1,0}(x;{1,1, -, 0, -,1})$ \cite[Eq. (1.125)]{mathai} and applying  the transformation $H_{p,q}^{u,v}\big[x \big|
\begin{array}{ccc} (a_i, k A_j)_p \\ (b_i, k B_j)_q \end{array}\big. \big]=\frac{1}{k}H_{p,q}^{u,v}\big[x^{\frac{1}{k}} \big|
\begin{array}{ccc} (a_i, A_j)_p \\ (b_i, B_j)_q \end{array}\big. \big]$.  Finally applying \cite[Eq. (2.3)]{mathai} yields the strongest-BS based coverage probability  as shwon in Proposition 3.

 \end{document}